\documentclass[aps,prl,twocolumn,showpacs]{revtex4-1}

\usepackage{graphicx}
\usepackage{dcolumn}
\usepackage{bm}
\usepackage{braket}
\usepackage{amsmath}
\usepackage{bbold}
\usepackage{verbatim}
\usepackage{hyperref}
\usepackage{dsfont}
\hypersetup{
    colorlinks=true,
    linkcolor=blue,
    filecolor=magenta,      
    urlcolor=cyan,
}

\begin{document}
\preprint{APS/123-QED}

\title[Further Compactifying Linear Optical Unitaries]{Further Compactifying Linear Optical Unitaries}

\author{B.A. Bell}
\email{b.bell@imperial.ac.uk}
\author{I.A. Walmsley}%
 
\affiliation{ 
Ultrafast Quantum Optics group, Department of Physics, Imperial College London,
London SW7 2AZ, UK
}%


\date{\today}

\begin{abstract}
Quantum integrated photonics requires large-scale linear optical circuitry, and for many applications it is desirable to have a universally programmable circuit, able to implement an arbitrary unitary transformation on a number of modes. This has been achieved using the Reck scheme, consisting of a network of Mach Zehnder interferometers containing a variable phase shifter in one path, as well as an external phase shifter after each Mach Zehnder. It subsequently became apparent that with symmetric Mach Zehnders containing a phase shift in both paths, the external phase shifts are redundant, resulting in a more compact circuit. The rectangular Clements scheme improves on the Reck scheme in terms of circuit depth, but it has been thought that an external phase-shifter was necessary after each Mach Zehnder. Here, we show that the Clements scheme can be realised using symmetric Mach Zehnders, requiring only a small number of external phase-shifters that do not contribute to the depth of the circuit. This will result in a significant saving in the length of these devices, allowing more complex circuits to fit onto a photonic chip, and reducing the propagation losses associated with these circuits. We also discuss how similar savings can be made to alternative schemes which have robustness to imbalanced beam-splitters.
\end{abstract}

\maketitle

\section{Introduction}

\begin{figure}
\includegraphics[width=\columnwidth]{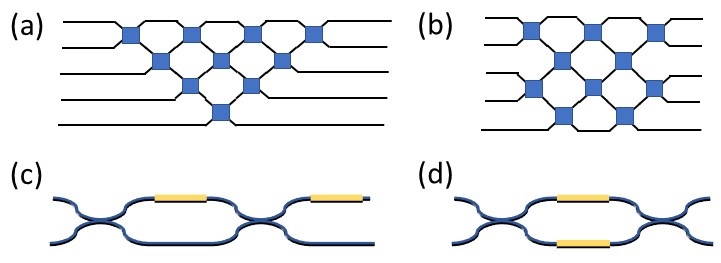}
\caption{\label{fig1} (a) A Reck scheme, where each blue square represents a tunable beamsplitter and a phase-shifter. (b) The Clements scheme. (c) The tunable beam-splitters can be realsied by MZIs with one internal phase-shifter. One external phase-shifter is needed to complete the unit cell. We term this an asymmetric MZI. (d) The symmetric MZI, with two internal phase shifts, makes for a more compact unit cell.}
\end{figure}

Optical quantum computing requires interferometric circuits to process quantum states of light~\cite{obrien07,Kok07}. In particular, integrated photonic circuits comprising networks of Mach Zehnder interferometers (MZIs) have emerged as a compact and versatile solution for realising reconfigurable linear optics, with applications to linear optical quantum computing~\cite{Carolan15, Metcalf14, Qiang18}, boson sampling~\cite{Spring13,Carolan14,Crespi13,Bell19}, high-dimensional encodings~\cite{Wang18,Taballione19}, quantum simulation~\cite{Sparrow18, Harris17}, photonic neural networks~\cite{Shen17}, and optical FPGAs~\cite{Zhuang15, Perez17}. Often, universal reconfigurability is essential or at least desirable, in the sense that a device can be programmed to realise any unitary transformation between the input and output modes. This can be achieved using the architecture of Reck et al. (the Reck scheme)~\cite{Reck94}, shown in Fig.~\ref{fig1}(a), a triangular mesh with each unit cell comprising a variable beam-splitter and a variable phase-shift. The rectangular architecture of Clements et al.~\cite{Clements16}, Fig.~\ref{fig1}(b), is also universal, and benefits from improved compactness as well as the balanced loss per channel, since each path has the same number of unit cells, that improves the overall fidelity of the circuit to the desired operation. In integrated photonics, the unit cell is usually implemented as an MZI with an internal phase-shifter in one of the arms, and an external phase-shifter on one output, as in Fig.~\ref{fig1}(c), which we will term an asymmetric MZI (aMZI). The internal phase-shifter controls the splitting ratio between the two outputs, while the external controls the relative phase between the outputs. The reason for using this primitive operation is that it is a universal 2x2 circuit.

Alternatively, Fig.~\ref{fig1}(d) shows a symmetric MZI (sMZI) with two internal and no external phase-shifters. This lacks control of the relative phase between outputs, yet it has been used in optical FPGAs~\cite{Zhuang15, Perez17} and it has been found that the sMZI can replace the aMZIs in a Reck scheme without compromising universality~\cite{Miller15, Ribeiro16}. The sMZI is attractive because it is more compact, without the need for an external phase-shifter, which can account for a significant fraction of the length of the circuit. A shorter structure not only occupies less area on a chip, but also suffers from less propagation loss. Further, there is a potential advantage in circuit loading from the controller drive signals, which benefits from the symmetric configuration in terms of e.g. heat distribution across the circuit. A rectangular architecture such as the Clements scheme but made up of sMZIs would be particularly beneficial for quantum integrated photonics, where large-scale circuits are required and transmission must be kept as high as possible. However sMZIs have not as yet been utilised in this way, because it is not clear that such a non-universal 2x2 element can generate a universal circuit, or that there is an efficient algorithm for finding the phases which will implement a target unitary.

In this work, we resolve this issue. We show that sMZIs can replace aMZIs in the Clements schemes, providing methods of matrix decomposition which generate the specific set of phases required to implement a given unitary matrix. We initially find that additional external phase-shifters are required in a diagonal layer through the middle of the circuit, as well as at a subset of the inputs and outputs. We subsequently show that the additional mid-circuit phase-shifts can be moved to otherwise vacant positions at the edge of the circuit, where they do not contribute to the overall length. This approximately halves the contribution of the phase-shifters to the length of a circuit, down to at most external phase-shifters at the inputs and outputs of the circuit, which are not required for some applications. We show how a similar advantage can be obtained for more general circuits comprising alternating layers of beam-splitters and phase-shifters, giving as an example the design of Fldzhyan et al.~\cite{Fldzhyan20}, which does not have a deterministic method of setting the phases but does appear to heuristically implement arbitrary unitary matrices with an improved level of robustness to imbalanced beam-splitters.

The remainder of this article is set out as follows. In Section II, we provide a matrix decomposition method for a Reck scheme made up of sMZIs. In Section III, we follow a near-identical method for the Clements scheme, showing how a layer of external phase-shifters arises mid-circuit. In Section IV, we propose a variant on the Clements scheme with additional phase-shifters at the edge of the circuit, which do not add to the overall length, and show that these can fulfil the same function as the external mid-circuit phase-shifts. In Section V we discuss how a similar technique can be applied to more general architectures, such as that of Fldzhyan et al.~\cite{Fldzhyan20}. In Section VI we summarise and draw conclusions.

\section{The Reck Scheme}

\begin{figure}
    \includegraphics[width=0.9\columnwidth]{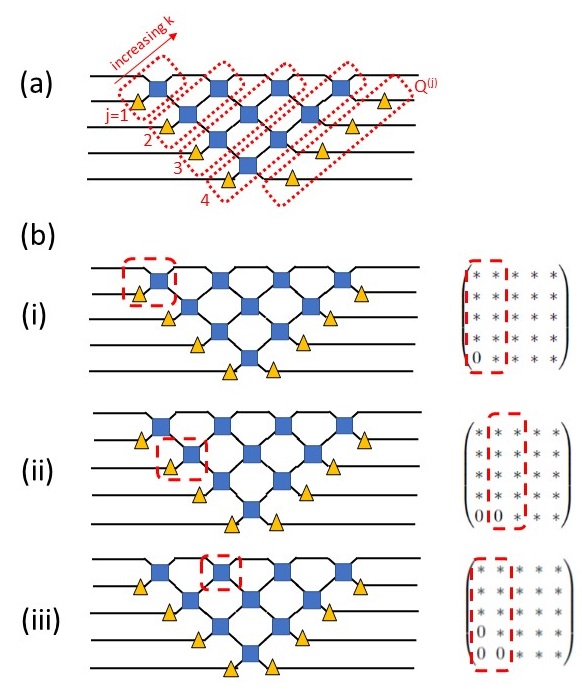}
    \caption{\label{fig2} (a) Layout of a Rech scheme with symmetric MZIs (blue squares) and external phase-shifters (yellow triangles). The elements are divided into diagonal rows labelled by $j$, within a row MZIs are labelled by $k$. (b) Steps (1) to (3) of the decomposition process. Circuit elements circled in red affect corresponding matrix elements circled in red, and their parameters are chosen to zero a succession of matrix elements.}
\end{figure}

We consider a Reck scheme~\cite{Reck94} where the unit cell is a sMZI, with additional external phase-shifts at the input and output of the entire circuit, on each mode except the first, as depicted in Fig.~\ref{fig2}(a). The matrix transformation implemented by one sMZI, applied to two adjacent modes, can be written as
\begin{equation}
    M=\begin{pmatrix}
    1~~ &&&&&\\ & \ddots &&&&\\
    && e^{i\Sigma}~\mathrm{sin}~\delta & e^{i\Sigma}~\mathrm{cos}~\delta &&\\
    && e^{i\Sigma}~\mathrm{cos}~\delta & -e^{i\Sigma}~\mathrm{sin}~\delta &&\\
    &&&& \ddots &\\&&&&& ~~1
    \end{pmatrix},
\end{equation}
where $\Sigma=(\theta_1+\theta_2)/2$, $\delta=(\theta_1-\theta_2)/2$, and $\theta_{1,2}$ are the values of the two internal phase-shifts. Meanwhile, an external phase-shift is given by
\begin{equation}
    P=\begin{pmatrix}
    1~~ &&&&\\ & \ddots &&&\\
    && e^{i\phi} &&\\
    &&& \ddots &\\&&&& ~~1
    \end{pmatrix}.
\end{equation}
The circuit is divided into diagonals, numbered $j=1$ to $m-1$, where $m$ is the number of modes in the circuit. The MZIs within each diagonal are numbered from bottom left to top right as $k=1$ to $j$. The MZI transformations within the circuit can be identified as $M^{(j,k)}$ with corresponding phase settings $\Sigma_{j,k}$ and $\delta_{j,k}$, while the input phase-shift operations associated with a diagonal are denoted $P^{(j)}$ with phase $\phi_j$. The output phase-shift operation applied to mode $j$ is labelled $Q^{(j)}$ with phase $\zeta_j$.

The matrix decomposition proceeds by applying the circuit operations to an auxillary matrix $V$ (initially set to $U^*$) so as to successively zero matrix elements. Fig.~\ref{fig2}(b) shows the first few operations used to zero elements. Multiplying $V$ from the right by an $M^{(j,k)}$ matrix mixes two columns of $V$ together - specifically the $y=j-k+1$ column is mixed with the $y+1$ column. A particular element $(x,y)$ can be set to zero by choosing $\delta_{j,k}$ such that:
\begin{equation}
    \mathrm{tan}~\delta_{j,k}=-V_{x,y}/V_{x,y+1}.
\end{equation}
This has a real solution for $\delta_{j,k}$ if $V_{x,y}$ and $V_{x,y+1}$ have the same complex phase, i.e. $\mathrm{arg}(V_{x,y})=\mathrm{arg}(V_{x,y+1})$. Since $\Sigma_{j,k}$ affects the phase of both columns equally, it cannot be chosen to achieve this condition - rather the phases need to be equalized by previous operations. The external phase-shifter $P^{(j)}$ can be used to match the phases for the first MZI in each diagonal, then for each MZI $\Sigma_{j,k}$ can be chosen to match the phases for the subsequent MZI. The detailed order of operations is given as follows:
\begin{enumerate}
\item Set an auxillary matrix $V\leftarrow U^*$.
\item For $j=1$ to $m-1$: \begin{itemize}
\item Set $x\leftarrow m$ and $y\leftarrow j$.
\item Set $V\leftarrow VP^{(j)}$, choosing $\phi_j=\mathrm{arg}(V_{x,y})-\mathrm{arg}(V_{x,y+1})$.
\item For $k=1$ to $j$:
\begin{itemize}
\item Set $V\leftarrow VM^{(j,k)}$, choosing $\delta_{j,k}$ such that $V_{x,y}$ is set to zero.
\item Choose $\Sigma_{j,k}$ such that $\mathrm{arg}(V_{x-1,y-1})=\mathrm{arg}(V_{x-1,y})$. For $k=j$, this choice is redundant.
\item Set $x\leftarrow x-1$ and $y\leftarrow y-1$.
\end{itemize}
\end{itemize}
\item For $j=2$ to $m$:
\begin{itemize}
    \item Set $V\leftarrow VQ^{(j)}$, choosing $\zeta_j=\mathrm{arg}(V_{1,1})-\mathrm{arg}(V_{j,j})$
\end{itemize}
\end{enumerate}
After Step 2, every element of $V$ below the diagonal has been set to zero. Since $V$ remains unitary, this implies that it is a diagonal matrix, where the remaining diagonal elements are complex with unit norm. Step 3 then applies the final phase-shifts such that $V$ is the identity matrix $\mathds{1}$, up to a global phase. Expanding $V$ as
\begin{equation}
    U^* P^{(1)}M^{(1,1)}...~ M^{(m-1,m-1)}Q^{(2)}...~Q^{(m)}=\mathds{1},
\end{equation}
it can be seen that
\begin{equation}
    U=Q^{(m)}...~ Q^{(1)}M^{(m-1,m-1)}...~ M^{(1,1)}P^{(1)},
\end{equation}
where we have used the fact that all of the circuit operations are symmetric unitaries, and so their complex conjugate is their inverse. Hence $U$ has been decomposed into the individual circuit operations.

The circuit consists of $\frac{1}{2}m(m-1)$ MZI, and so $m(m-1)$ internal phase-shifters. In each row of MZI, the final choice of $\Sigma_{j,j}$ is redundant, and so in these MZI one of the internal phase-shifters could be omitted. This would leave $(m-1)^2$ internal phase-shifters and $2(m-1)$ external. This is equal to the number of free parameters in an $m\times m$ unitary matrix, and hence is the minimum required. We note that in situations where the output modes are to be connected directly to phase-insensitive detectors, the external phase-shifts at the output are redundant and could be omitted. In some situations, the external phases at the inputs could also be redundant, for instance if single photon input states are used.

\section{The Clements Scheme}

\begin{figure}
    \includegraphics[width=0.8\columnwidth]{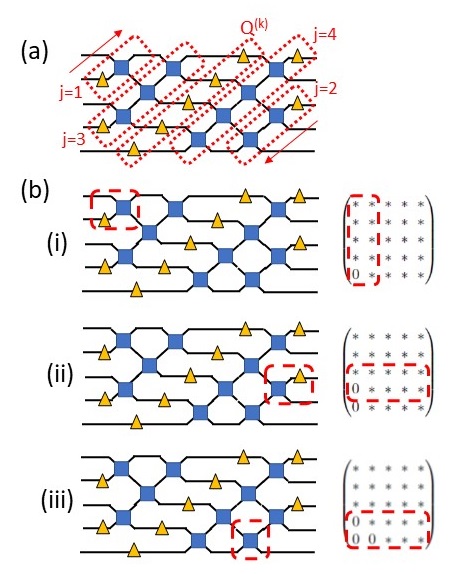}
    \caption{\label{fig3} (a) Labelling of diagonal rows of MZI and ordering within a row for the Clements scheme. For even rows, the ordering is reversed, beginning from the outputs. (b) Choosing the phase-shifts in the first two rows. (i) applies the MZI in the 1st row by multiplying the auxillary matrix from the right, mixing the first two columns together to zero the bottom left element. (ii) and (iii) implement the MZIs in the 2nd row to zero further elements, by multiplying the auxillary matrix from the left. This has the effect of mixing two rows instead of two columns. (c) The residual phases are left mid-circuit rather than at the outputs as with the Reck scheme. However they can be absorbed into the action of surrounding phase-shifters with minimal changes to the circuit.}
\end{figure}

We now consider the Clements scheme~\cite{Clements16} circuit shown in Fig.~\ref{fig3}(a). The MZI are organised into diagonals labelled by $j$ as in the previous section, but for even $j$ the direction of $k$ has been reversed - as a result, the associated $P^{(j)}$ phase-shifters have been moved from input to output. In the decomposition, the order in which matrix elements are zeroed within an even $j$ diagonal is also reversed, and the corresponding circuit operations are applied to $V$ by left multiplication rather than right multiplication. As a result these operations mix adjacent rows of the $V$ matrix rather than adjacent columns - the first few steps of this process are shown in Fig.~\ref{fig3}(b). The $Q^{(j)}$ operations appear in a diagonal line through the middle of the circuit, which is not a usual feature of the Clements scheme - in the next section we will show how these phase-shifts can be relocated to positions where they do not add to the overall length. In other respects the decomposition is identical to that of the Reck scheme.
\begin{enumerate}
    \item Set an auxillary matrix $V\leftarrow U^*$.
    \item For $j=1$ to $m-1$:
    \begin{itemize}
        \item if $j$ is odd:
        \begin{itemize}
            \item Set $x\leftarrow m$ and $y\leftarrow j$.
            \item Set $V\leftarrow VP^{(j)}$, choosing $\phi_j=\mathrm{arg}(V_{x,y})-\mathrm{arg}(V_{x,y+1})$.
            \item For $k=1$ to $j$:
            \begin{itemize}
                \item Set $V\leftarrow V M^{(j,k)}$, choosing $\delta_{j,k}$ to zero $V_{x,y}$.
                \item Choose $\Sigma_{j,k}$ so that $\mathrm{arg}(V_{x-1,y-1})=\mathrm{arg}(V_{x-1,y})$.
                \item Set $x\leftarrow x-1$ and $y\leftarrow y-1$.
            \end{itemize}
        \end{itemize}
        \item if $j$ is even:
        \begin{itemize}
            \item Set $x\leftarrow m-j+1$ and $y\leftarrow 1$.
            \item Set $V\leftarrow P^{(j)}V$, choosing $\phi^{(j)}=\mathrm{arg}(V_{x,y})-\mathrm{arg}(V_{x-1,y})$.
            \item For $k=1$ to $j$:
            \begin{itemize}
                \item Set $V\leftarrow M^{(j,k)}V$, choosing $\delta_{j,k}$ to zero $V_{x,y}$.
                \item Choose $\Sigma_{j,k}$ so that $\mathrm{arg}(V_{x+1,y+1})=\mathrm{arg}(V_{x,y+1})$.
                \item Set $x\leftarrow x+1$ and $y\leftarrow y+1$.
            \end{itemize}
        \end{itemize}
    \end{itemize}
    \item For $j=2$ to $m$:
    \begin{itemize}
        \item Set $V\leftarrow VQ^{(j)}$, choosing $\zeta_j=\mathrm{arg}(V_{1,1})-\mathrm{arg}(V_{j,j})$.
    \end{itemize}
\end{enumerate}
As before, at the end of Step 3, $V$ is the identity matrix up to a global phase. However, now when expanding $V$:
\begin{equation}
    \begin{split}
        ...M^{(4,4)}...M^{(2,1)}P^{(2)}~U^*~ P^{(1)}M^{(1,1)}...M^{(3,3)}...&~Q^{(2)}...Q^{(m)}\\ &=\mathds{1},
    \end{split}
\end{equation}
i.e. the operations with even $j$ are to the left of $U^*$ rather than the right. It can be seen that this naturally places the $Q^{(j)}$ operations which were applied last in the middle of the circuit:
\begin{equation}
    U=P^{(2)}M^{(2,1)}...M^{(4,4)}...Q^{(2)}...Q^{(m)}...M^{(3,3)}...M^{(1,1)}P^{(1)}.
\end{equation}

\section{Relocating residual phase-shifts}

\begin{figure}
    \includegraphics[width=0.8\columnwidth]{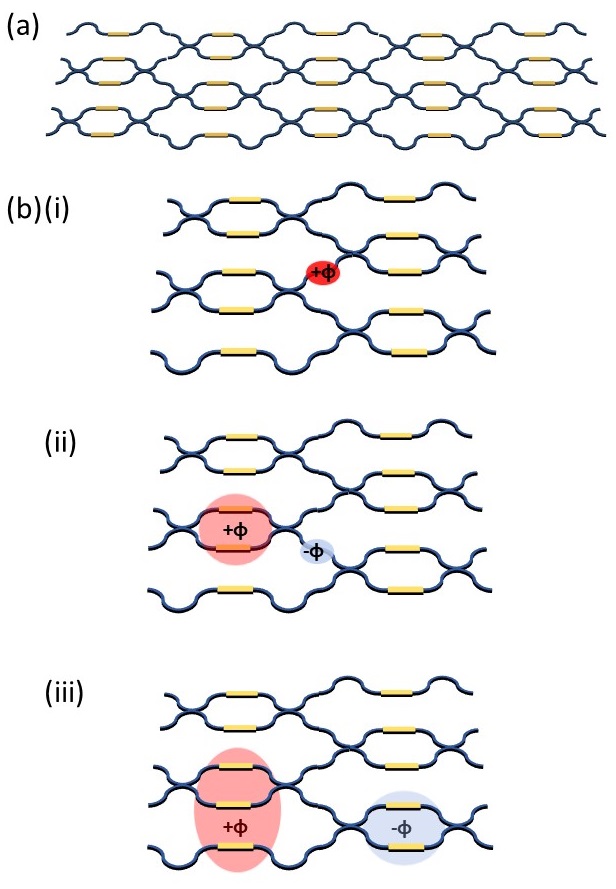}
    \caption{\label{fig4} (a) A Clements scheme circuit consisting of symmetric MZIs, with additional phase-shifters on the waveguides at the edges of the circuit. (b)(i) A desired phase-shift $\phi$ between layers of MZIs is implemented by (ii) adding $+\phi$ to both phases in an adjacents MZI; this moves the required phase to one of opposite sign on the mode immediately below the desired target phase. (iii) Repeating this operation until the residual phase is at the edge of the circuit, it can then be implemented by one of the additional phase-shifters.}
\end{figure}

We now propose a variant on the Clements scheme as shown in Fig.~\ref{fig4}(a), consisting of a rectangular network of sMZI. For each vertical layer  of MZIs, there is one odd mode at the edge which is not involved in a MZI (for an even total number of modes, alternating layers have zero or two modes not involved in a MZI). The only change compared to the circuit considered in the previous section is that tunable phase-shifters are added to these sections of waveguide, instead of placing the $Q^{(k)}$ phase-shifters in the middle of the circuit. The new phase-shifters do not add to the overall length of the circuit, since path-length matching dictates that these sections of waveguide are anyway as long as if they had been involved in a MZI. The external phase-shifters at the inputs and outputs are not shown here.

Now, if a phase-shift $\phi$ is required on a single waveguide between layers of MZI, we follow a new procedure, shown in Fig.~\ref{fig4}(b). Adding $+\phi$ to both phase-shifters in the MZI to the left implements the phase, but a $-\phi$ shift is now required to correct the effect on the waveguide below the original phase-shift. Now an MZI to the right can be used to implement the $-\phi$, while requiring a $+\phi$ correction to a lower waveguide. This can be repeated until the residual phase-shift is moved to the edge of the circuit, where it can be implemented directly with one of the new phase-shifters. This relies on the fact that the operation of applying the same phase-shift to two modes will commute with a beam-splitter operation on those two modes.

This demonstrates that the new circuit is universal - one can follow the decomposition given in the previous section and then implement the $Q^{(k)}$ phase-shifts using this method. Two of these operations are already at the edges of the circuit, and so can trivially be implemented by the edge phase-shifters. One could also follow the original Clements decomposition intended for aMZIs, and then implement all of the external phase-shifts required by this method of absorbing them into surrounding sMZIs and shifting them to the edge of the circuit.

For each layer of MZIs, there are now $m$ phase-shifters. Since adding the same phase to all the phase-shifters in a layer only adds a global phase, there is redundancy here, and any one phase-shifter could be removed from each layer. The circuit would then have the minimum number of degrees of freedom required to implement an arbitrary unitary. On the other hand, this is quite a healthy redundancy to have, since any one phase-shifter in each layer can fail (e.g. due to fabrication error), and the overall operation of the circuit will be preserved.

\section{Error Tolerant Designs}

The Reck and Clements schemes are provably universal on the assumption that every beam-splitter in the circuit has an exactly balanced splitting ratio. If some beam-splitters deviate from this, for instance due to uncertainties in fabrication, then the MZIs no longer have full tunability in their splitting ratio, and some unitary transformations become inaccessible to the circuit. Several alternative designs have been proposed which show improved robustness to randomized beam-splitters, including adding redundant layers of MZI~\cite{Burgwal17}, adding permutations of waveguides between MZI layers~\cite{Pai19}, and the design of Fldzhyan et al.~\cite{Fldzhyan20}, which makes use of alternating layers of beam-splitters and phase-shifters in an arrangement which does not map onto a network of MZIs. For these designs, there is generally no known deterministic method of decomposing them into elementary 2x2 unitaries. Rather the phase settings are optimised to minimise the distance to some target unitary matrix.

\begin{figure}
\includegraphics[width=\columnwidth]{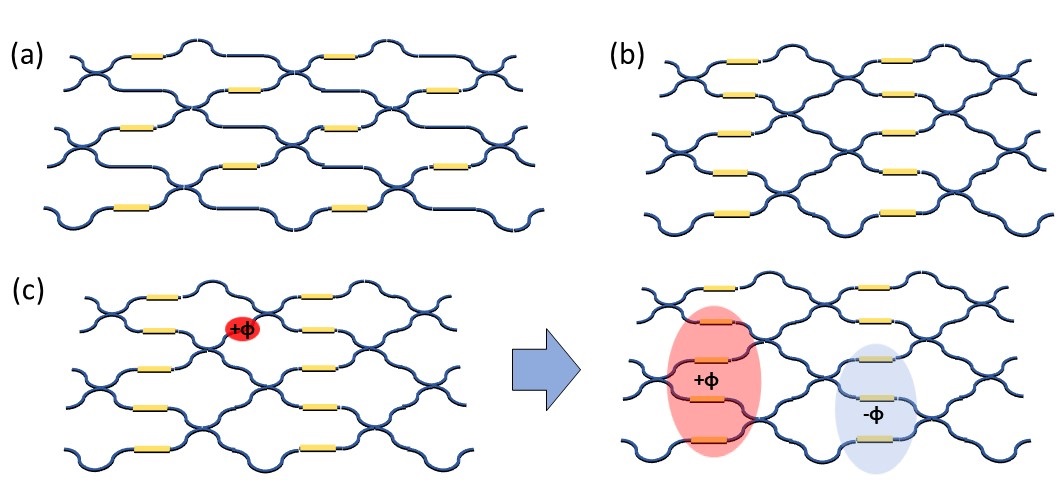}
\caption{\label{fig5} (a) The Fldzhyan et al. design, consisting of alternating layers of beam-splitters and phase-shifters. (b) An equivalent design where every other layer of phase-shifters have been moved to the preceding layer. (c) Any required phase $\phi$ in a missing layer can be implemented by applying $\pm\phi$ shifts in neighbouring layers.}
\end{figure}

For rectangular meshes of MZI similar to the Clements scheme it is fairly clear where aMZI can be replaced by sMZI, hence we focus on the Fldzhyan design here. This design is appealing because the depth of the circuit and the number of elements is identical to the Clements scheme, while the robustness to beam-splitter imbalance is improved for Haar-random target unitaries. The circuit layout is as shown in Fig.~\ref{fig5}(a) - here, only four layers of beam-splitters and phase-shifters are shown, with $2m$ such layers required for universality. Fig.~\ref{fig5}(b) shows a compactified design, where every other layer of phase-shifters has been moved into the preceding layer. It can be seen that this design is equivalent to the original one, since where-ever a phase is required in a removed layer, it can be applied by adjusting phases in the neighbouring layers. Fig.~\ref{fig5}(c) shows how a phase $\phi$ in a removed layer is implemented by applying $+\phi$ to a subset of phase-shifters in the neighbouring layer to the left, and $-\phi$ to a subset in the neighbouring layer to the right. As previously, this relies on the fact that equal phase-shifts applied to both two modes involved in a beam-splitter can be commuted to the input or output of the beam-splitter. This remains valid regardless of the splitting ratio of the beam-splitter.

\section{Conclusions}

In summary, we have shown that the aMZI can be replaced by sMZI in a Clements style rectangular network while retaining universality, providing a deterministic and efficient method of selecting the phases required to implement a target unitary. We expect that sMZI can replace aMZI in related designs making use of a rectangular structure, and have used similar logic to suggest a more compact but equivalent form of the circuit of Fldzhyan et al.\cite{Fldzhyan20}. Using sMZIs implies that the circuit length taken up by phase-shifters is approximately halved, compared to aMZIs. This can be a significant fraction of the overall circuit length in many integrated photonic platforms - for example in silicon-on-insulator~\cite{Shen17}, silica-on-silicon~\cite{Mennea18}, and lithium niobate-on-insulator~\cite{Boes18}, the length of a thermo-optic phase shifter is comparable to that of a beam-splitter, whereas in silicon nitride the phase-shifters are relatively long and can be the dominant contribution to the circuit length~\cite{Taballione20,Arrazola21}. We expect that this saving in length will allow larger universal linear optical circuits to fit on a chip, while reducing the propagation loss, helping to realise large-scale and complex quantum photonic computation and simulation. 

We acknowledge funding from the EPSRC UK Quantum Technologies Programme (EP/T001062/1) via the Quantum Computing and Simulation hub. Bryn Bell is supported by a European Commission Marie Skłodowska Curie Individual Fellowship (FrEQuMP, 846073).

\section*{Data Availability}

The data that support the findings of this study are available from the corresponding author upon reasonable request.


\bibliography{aipsamp}

\providecommand{\noopsort}[1]{}\providecommand{\singleletter}[1]{#1}%
\begin{thebibliography}{27}%
\makeatletter
\providecommand \@ifxundefined [1]{%
 \@ifx{#1\undefined}
}%
\providecommand \@ifnum [1]{%
 \ifnum #1\expandafter \@firstoftwo
 \else \expandafter \@secondoftwo
 \fi
}%
\providecommand \@ifx [1]{%
 \ifx #1\expandafter \@firstoftwo
 \else \expandafter \@secondoftwo
 \fi
}%
\providecommand \natexlab [1]{#1}%
\providecommand \enquote  [1]{``#1''}%
\providecommand \bibnamefont  [1]{#1}%
\providecommand \bibfnamefont [1]{#1}%
\providecommand \citenamefont [1]{#1}%
\providecommand \href@noop [0]{\@secondoftwo}%
\providecommand \href [0]{\begingroup \@sanitize@url \@href}%
\providecommand \@href[1]{\@@startlink{#1}\@@href}%
\providecommand \@@href[1]{\endgroup#1\@@endlink}%
\providecommand \@sanitize@url [0]{\catcode `\\12\catcode `\$12\catcode
  `\&12\catcode `\#12\catcode `\^12\catcode `\_12\catcode `\%12\relax}%
\providecommand \@@startlink[1]{}%
\providecommand \@@endlink[0]{}%
\providecommand \url  [0]{\begingroup\@sanitize@url \@url }%
\providecommand \@url [1]{\endgroup\@href {#1}{\urlprefix }}%
\providecommand \urlprefix  [0]{URL }%
\providecommand \Eprint [0]{\href }%
\providecommand \doibase [0]{http://dx.doi.org/}%
\providecommand \selectlanguage [0]{\@gobble}%
\providecommand \bibinfo  [0]{\@secondoftwo}%
\providecommand \bibfield  [0]{\@secondoftwo}%
\providecommand \translation [1]{[#1]}%
\providecommand \BibitemOpen [0]{}%
\providecommand \bibitemStop [0]{}%
\providecommand \bibitemNoStop [0]{.\EOS\space}%
\providecommand \EOS [0]{\spacefactor3000\relax}%
\providecommand \BibitemShut  [1]{\csname bibitem#1\endcsname}%
\let\auto@bib@innerbib\@empty
\bibitem [{\citenamefont {O{\textquoteright}Brien}(2007)}]{obrien07}%
  \BibitemOpen
  \bibfield  {author} {\bibinfo {author} {\bibfnamefont {J.~L.}\ \bibnamefont
  {O{\textquoteright}Brien}},\ }\href {\doibase 10.1126/science.1142892}
  {\bibfield  {journal} {\bibinfo  {journal} {Science}\ }\textbf {\bibinfo
  {volume} {318}},\ \bibinfo {pages} {1567} (\bibinfo {year}
  {2007})}\BibitemShut {NoStop}%
\bibitem [{\citenamefont {Kok}\ \emph {et~al.}(2007)\citenamefont {Kok},
  \citenamefont {Munro}, \citenamefont {Nemoto}, \citenamefont {Ralph},
  \citenamefont {Dowling},\ and\ \citenamefont {Milburn}}]{Kok07}%
  \BibitemOpen
  \bibfield  {author} {\bibinfo {author} {\bibfnamefont {P.}~\bibnamefont
  {Kok}}, \bibinfo {author} {\bibfnamefont {W.~J.}\ \bibnamefont {Munro}},
  \bibinfo {author} {\bibfnamefont {K.}~\bibnamefont {Nemoto}}, \bibinfo
  {author} {\bibfnamefont {T.~C.}\ \bibnamefont {Ralph}}, \bibinfo {author}
  {\bibfnamefont {J.~P.}\ \bibnamefont {Dowling}}, \ and\ \bibinfo {author}
  {\bibfnamefont {G.~J.}\ \bibnamefont {Milburn}},\ }\href {\doibase
  10.1103/RevModPhys.79.135} {\bibfield  {journal} {\bibinfo  {journal} {Rev.
  Mod. Phys.}\ }\textbf {\bibinfo {volume} {79}},\ \bibinfo {pages} {135}
  (\bibinfo {year} {2007})}\BibitemShut {NoStop}%
\bibitem [{\citenamefont {Carolan}\ \emph {et~al.}(2015)\citenamefont
  {Carolan}, \citenamefont {Harrold}, \citenamefont {Sparrow}, \citenamefont
  {Mart{\'\i}n-L{\'o}pez}, \citenamefont {Russell}, \citenamefont
  {Silverstone}, \citenamefont {Shadbolt}, \citenamefont {Matsuda},
  \citenamefont {Oguma}, \citenamefont {Itoh}, \citenamefont {Marshall},
  \citenamefont {Thompson}, \citenamefont {Matthews}, \citenamefont
  {Hashimoto}, \citenamefont {O{\textquoteright}Brien},\ and\ \citenamefont
  {Laing}}]{Carolan15}%
  \BibitemOpen
  \bibfield  {author} {\bibinfo {author} {\bibfnamefont {J.}~\bibnamefont
  {Carolan}}, \bibinfo {author} {\bibfnamefont {C.}~\bibnamefont {Harrold}},
  \bibinfo {author} {\bibfnamefont {C.}~\bibnamefont {Sparrow}}, \bibinfo
  {author} {\bibfnamefont {E.}~\bibnamefont {Mart{\'\i}n-L{\'o}pez}}, \bibinfo
  {author} {\bibfnamefont {N.~J.}\ \bibnamefont {Russell}}, \bibinfo {author}
  {\bibfnamefont {J.~W.}\ \bibnamefont {Silverstone}}, \bibinfo {author}
  {\bibfnamefont {P.~J.}\ \bibnamefont {Shadbolt}}, \bibinfo {author}
  {\bibfnamefont {N.}~\bibnamefont {Matsuda}}, \bibinfo {author} {\bibfnamefont
  {M.}~\bibnamefont {Oguma}}, \bibinfo {author} {\bibfnamefont
  {M.}~\bibnamefont {Itoh}}, \bibinfo {author} {\bibfnamefont {G.~D.}\
  \bibnamefont {Marshall}}, \bibinfo {author} {\bibfnamefont {M.~G.}\
  \bibnamefont {Thompson}}, \bibinfo {author} {\bibfnamefont {J.~C.~F.}\
  \bibnamefont {Matthews}}, \bibinfo {author} {\bibfnamefont {T.}~\bibnamefont
  {Hashimoto}}, \bibinfo {author} {\bibfnamefont {J.~L.}\ \bibnamefont
  {O{\textquoteright}Brien}}, \ and\ \bibinfo {author} {\bibfnamefont
  {A.}~\bibnamefont {Laing}},\ }\href {\doibase 10.1126/science.aab3642}
  {\bibfield  {journal} {\bibinfo  {journal} {Science}\ }\textbf {\bibinfo
  {volume} {349}},\ \bibinfo {pages} {711} (\bibinfo {year}
  {2015})}\BibitemShut {NoStop}%
\bibitem [{\citenamefont {Metcalf}\ \emph {et~al.}(2014)\citenamefont
  {Metcalf}, \citenamefont {Spring}, \citenamefont {Humphreys}, \citenamefont
  {Thomas-Peter}, \citenamefont {Barbieri}, \citenamefont {Kolthammer},
  \citenamefont {Jin}, \citenamefont {Langford}, \citenamefont {Kundys},
  \citenamefont {Gates}, \citenamefont {Smith}, \citenamefont {Smith},\ and\
  \citenamefont {Walmsley}}]{Metcalf14}%
  \BibitemOpen
  \bibfield  {author} {\bibinfo {author} {\bibfnamefont {B.~J.}\ \bibnamefont
  {Metcalf}}, \bibinfo {author} {\bibfnamefont {J.~B.}\ \bibnamefont {Spring}},
  \bibinfo {author} {\bibfnamefont {P.~C.}\ \bibnamefont {Humphreys}}, \bibinfo
  {author} {\bibfnamefont {N.}~\bibnamefont {Thomas-Peter}}, \bibinfo {author}
  {\bibfnamefont {M.}~\bibnamefont {Barbieri}}, \bibinfo {author}
  {\bibfnamefont {W.~S.}\ \bibnamefont {Kolthammer}}, \bibinfo {author}
  {\bibfnamefont {X.-M.}\ \bibnamefont {Jin}}, \bibinfo {author} {\bibfnamefont
  {N.~K.}\ \bibnamefont {Langford}}, \bibinfo {author} {\bibfnamefont
  {D.}~\bibnamefont {Kundys}}, \bibinfo {author} {\bibfnamefont {J.~C.}\
  \bibnamefont {Gates}}, \bibinfo {author} {\bibfnamefont {B.~J.}\ \bibnamefont
  {Smith}}, \bibinfo {author} {\bibfnamefont {P.~G.~R.}\ \bibnamefont {Smith}},
  \ and\ \bibinfo {author} {\bibfnamefont {I.~A.}\ \bibnamefont {Walmsley}},\
  }\href {\doibase 10.1038/nphoton.2014.217} {\bibfield  {journal} {\bibinfo
  {journal} {Nature Photonics}\ }\textbf {\bibinfo {volume} {8}},\ \bibinfo
  {pages} {770} (\bibinfo {year} {2014})}\BibitemShut {NoStop}%
\bibitem [{\citenamefont {Qiang}\ \emph {et~al.}(2018)\citenamefont {Qiang},
  \citenamefont {Zhou}, \citenamefont {Wang}, \citenamefont {Wilkes},
  \citenamefont {Loke}, \citenamefont {O'Gara}, \citenamefont {Kling},
  \citenamefont {Marshall}, \citenamefont {Santagati}, \citenamefont {Ralph},
  \citenamefont {Wang}, \citenamefont {O'Brien}, \citenamefont {Thompson},\
  and\ \citenamefont {Matthews}}]{Qiang18}%
  \BibitemOpen
  \bibfield  {author} {\bibinfo {author} {\bibfnamefont {X.}~\bibnamefont
  {Qiang}}, \bibinfo {author} {\bibfnamefont {X.}~\bibnamefont {Zhou}},
  \bibinfo {author} {\bibfnamefont {J.}~\bibnamefont {Wang}}, \bibinfo {author}
  {\bibfnamefont {C.~M.}\ \bibnamefont {Wilkes}}, \bibinfo {author}
  {\bibfnamefont {T.}~\bibnamefont {Loke}}, \bibinfo {author} {\bibfnamefont
  {S.}~\bibnamefont {O'Gara}}, \bibinfo {author} {\bibfnamefont
  {L.}~\bibnamefont {Kling}}, \bibinfo {author} {\bibfnamefont {G.~D.}\
  \bibnamefont {Marshall}}, \bibinfo {author} {\bibfnamefont {R.}~\bibnamefont
  {Santagati}}, \bibinfo {author} {\bibfnamefont {T.~C.}\ \bibnamefont
  {Ralph}}, \bibinfo {author} {\bibfnamefont {J.~B.}\ \bibnamefont {Wang}},
  \bibinfo {author} {\bibfnamefont {J.~L.}\ \bibnamefont {O'Brien}}, \bibinfo
  {author} {\bibfnamefont {M.~G.}\ \bibnamefont {Thompson}}, \ and\ \bibinfo
  {author} {\bibfnamefont {J.~C.~F.}\ \bibnamefont {Matthews}},\ }\href
  {\doibase 10.1038/s41566-018-0236-y} {\bibfield  {journal} {\bibinfo
  {journal} {Nature Photonics}\ }\textbf {\bibinfo {volume} {12}},\ \bibinfo
  {pages} {534} (\bibinfo {year} {2018})}\BibitemShut {NoStop}%
\bibitem [{\citenamefont {Spring}\ \emph {et~al.}(2013)\citenamefont {Spring},
  \citenamefont {Metcalf}, \citenamefont {Humphreys}, \citenamefont
  {Kolthammer}, \citenamefont {Jin}, \citenamefont {Barbieri}, \citenamefont
  {Datta}, \citenamefont {Thomas-Peter}, \citenamefont {Langford},
  \citenamefont {Kundys}, \citenamefont {Gates}, \citenamefont {Smith},
  \citenamefont {Smith},\ and\ \citenamefont {Walmsley}}]{Spring13}%
  \BibitemOpen
  \bibfield  {author} {\bibinfo {author} {\bibfnamefont {J.~B.}\ \bibnamefont
  {Spring}}, \bibinfo {author} {\bibfnamefont {B.~J.}\ \bibnamefont {Metcalf}},
  \bibinfo {author} {\bibfnamefont {P.~C.}\ \bibnamefont {Humphreys}}, \bibinfo
  {author} {\bibfnamefont {W.~S.}\ \bibnamefont {Kolthammer}}, \bibinfo
  {author} {\bibfnamefont {X.-M.}\ \bibnamefont {Jin}}, \bibinfo {author}
  {\bibfnamefont {M.}~\bibnamefont {Barbieri}}, \bibinfo {author}
  {\bibfnamefont {A.}~\bibnamefont {Datta}}, \bibinfo {author} {\bibfnamefont
  {N.}~\bibnamefont {Thomas-Peter}}, \bibinfo {author} {\bibfnamefont {N.~K.}\
  \bibnamefont {Langford}}, \bibinfo {author} {\bibfnamefont {D.}~\bibnamefont
  {Kundys}}, \bibinfo {author} {\bibfnamefont {J.~C.}\ \bibnamefont {Gates}},
  \bibinfo {author} {\bibfnamefont {B.~J.}\ \bibnamefont {Smith}}, \bibinfo
  {author} {\bibfnamefont {P.~G.~R.}\ \bibnamefont {Smith}}, \ and\ \bibinfo
  {author} {\bibfnamefont {I.~A.}\ \bibnamefont {Walmsley}},\ }\href {\doibase
  10.1126/science.1231692} {\bibfield  {journal} {\bibinfo  {journal}
  {Science}\ }\textbf {\bibinfo {volume} {339}},\ \bibinfo {pages} {798}
  (\bibinfo {year} {2013})}\BibitemShut {NoStop}%
\bibitem [{\citenamefont {Carolan}\ \emph {et~al.}(2014)\citenamefont
  {Carolan}, \citenamefont {Meinecke}, \citenamefont {Shadbolt}, \citenamefont
  {Russell}, \citenamefont {Ismail}, \citenamefont {W{\"o}rhoff}, \citenamefont
  {Rudolph}, \citenamefont {Thompson}, \citenamefont {O'Brien}, \citenamefont
  {Matthews},\ and\ \citenamefont {Laing}}]{Carolan14}%
  \BibitemOpen
  \bibfield  {author} {\bibinfo {author} {\bibfnamefont {J.}~\bibnamefont
  {Carolan}}, \bibinfo {author} {\bibfnamefont {J.~D.~A.}\ \bibnamefont
  {Meinecke}}, \bibinfo {author} {\bibfnamefont {P.~J.}\ \bibnamefont
  {Shadbolt}}, \bibinfo {author} {\bibfnamefont {N.~J.}\ \bibnamefont
  {Russell}}, \bibinfo {author} {\bibfnamefont {N.}~\bibnamefont {Ismail}},
  \bibinfo {author} {\bibfnamefont {K.}~\bibnamefont {W{\"o}rhoff}}, \bibinfo
  {author} {\bibfnamefont {T.}~\bibnamefont {Rudolph}}, \bibinfo {author}
  {\bibfnamefont {M.~G.}\ \bibnamefont {Thompson}}, \bibinfo {author}
  {\bibfnamefont {J.~L.}\ \bibnamefont {O'Brien}}, \bibinfo {author}
  {\bibfnamefont {J.~C.~F.}\ \bibnamefont {Matthews}}, \ and\ \bibinfo {author}
  {\bibfnamefont {A.}~\bibnamefont {Laing}},\ }\href {\doibase
  10.1038/nphoton.2014.152} {\bibfield  {journal} {\bibinfo  {journal} {Nature
  Photonics}\ }\textbf {\bibinfo {volume} {8}},\ \bibinfo {pages} {621–626}
  (\bibinfo {year} {2014})}\BibitemShut {NoStop}%
\bibitem [{\citenamefont {Crespi}\ \emph {et~al.}(2013)\citenamefont {Crespi},
  \citenamefont {Osellame}, \citenamefont {Ramponi}, \citenamefont {Brod},
  \citenamefont {Galv{\~a}o}, \citenamefont {Spagnolo}, \citenamefont
  {Vitelli}, \citenamefont {Maiorino}, \citenamefont {Mataloni},\ and\
  \citenamefont {Sciarrino}}]{Crespi13}%
  \BibitemOpen
  \bibfield  {author} {\bibinfo {author} {\bibfnamefont {A.}~\bibnamefont
  {Crespi}}, \bibinfo {author} {\bibfnamefont {R.}~\bibnamefont {Osellame}},
  \bibinfo {author} {\bibfnamefont {R.}~\bibnamefont {Ramponi}}, \bibinfo
  {author} {\bibfnamefont {D.~J.}\ \bibnamefont {Brod}}, \bibinfo {author}
  {\bibfnamefont {E.~F.}\ \bibnamefont {Galv{\~a}o}}, \bibinfo {author}
  {\bibfnamefont {N.}~\bibnamefont {Spagnolo}}, \bibinfo {author}
  {\bibfnamefont {C.}~\bibnamefont {Vitelli}}, \bibinfo {author} {\bibfnamefont
  {E.}~\bibnamefont {Maiorino}}, \bibinfo {author} {\bibfnamefont
  {P.}~\bibnamefont {Mataloni}}, \ and\ \bibinfo {author} {\bibfnamefont
  {F.}~\bibnamefont {Sciarrino}},\ }\href {\doibase 10.1038/nphoton.2013.112}
  {\bibfield  {journal} {\bibinfo  {journal} {Nature Photonics}\ }\textbf
  {\bibinfo {volume} {7}},\ \bibinfo {pages} {545} (\bibinfo {year}
  {2013})}\BibitemShut {NoStop}%
\bibitem [{\citenamefont {Bell}\ \emph {et~al.}(2019)\citenamefont {Bell},
  \citenamefont {Thekkadath}, \citenamefont {Ge}, \citenamefont {Cai},\ and\
  \citenamefont {Walmsley}}]{Bell19}%
  \BibitemOpen
  \bibfield  {author} {\bibinfo {author} {\bibfnamefont {B.~A.}\ \bibnamefont
  {Bell}}, \bibinfo {author} {\bibfnamefont {G.~S.}\ \bibnamefont
  {Thekkadath}}, \bibinfo {author} {\bibfnamefont {R.}~\bibnamefont {Ge}},
  \bibinfo {author} {\bibfnamefont {X.}~\bibnamefont {Cai}}, \ and\ \bibinfo
  {author} {\bibfnamefont {I.~A.}\ \bibnamefont {Walmsley}},\ }\href {\doibase
  10.1364/OE.27.035646} {\bibfield  {journal} {\bibinfo  {journal} {Opt.
  Express}\ }\textbf {\bibinfo {volume} {27}},\ \bibinfo {pages} {35646}
  (\bibinfo {year} {2019})}\BibitemShut {NoStop}%
\bibitem [{\citenamefont {Wang}\ \emph {et~al.}(2018)\citenamefont {Wang},
  \citenamefont {Paesani}, \citenamefont {Ding}, \citenamefont {Santagati},
  \citenamefont {Skrzypczyk}, \citenamefont {Salavrakos}, \citenamefont {Tura},
  \citenamefont {Augusiak}, \citenamefont {Man{\v c}inska}, \citenamefont
  {Bacco}, \citenamefont {Bonneau}, \citenamefont {Silverstone}, \citenamefont
  {Gong}, \citenamefont {Ac{\'\i}n}, \citenamefont {Rottwitt}, \citenamefont
  {Oxenl{\o}we}, \citenamefont {O{\textquoteright}Brien}, \citenamefont
  {Laing},\ and\ \citenamefont {Thompson}}]{Wang18}%
  \BibitemOpen
  \bibfield  {author} {\bibinfo {author} {\bibfnamefont {J.}~\bibnamefont
  {Wang}}, \bibinfo {author} {\bibfnamefont {S.}~\bibnamefont {Paesani}},
  \bibinfo {author} {\bibfnamefont {Y.}~\bibnamefont {Ding}}, \bibinfo {author}
  {\bibfnamefont {R.}~\bibnamefont {Santagati}}, \bibinfo {author}
  {\bibfnamefont {P.}~\bibnamefont {Skrzypczyk}}, \bibinfo {author}
  {\bibfnamefont {A.}~\bibnamefont {Salavrakos}}, \bibinfo {author}
  {\bibfnamefont {J.}~\bibnamefont {Tura}}, \bibinfo {author} {\bibfnamefont
  {R.}~\bibnamefont {Augusiak}}, \bibinfo {author} {\bibfnamefont
  {L.}~\bibnamefont {Man{\v c}inska}}, \bibinfo {author} {\bibfnamefont
  {D.}~\bibnamefont {Bacco}}, \bibinfo {author} {\bibfnamefont
  {D.}~\bibnamefont {Bonneau}}, \bibinfo {author} {\bibfnamefont {J.~W.}\
  \bibnamefont {Silverstone}}, \bibinfo {author} {\bibfnamefont
  {Q.}~\bibnamefont {Gong}}, \bibinfo {author} {\bibfnamefont {A.}~\bibnamefont
  {Ac{\'\i}n}}, \bibinfo {author} {\bibfnamefont {K.}~\bibnamefont {Rottwitt}},
  \bibinfo {author} {\bibfnamefont {L.~K.}\ \bibnamefont {Oxenl{\o}we}},
  \bibinfo {author} {\bibfnamefont {J.~L.}\ \bibnamefont
  {O{\textquoteright}Brien}}, \bibinfo {author} {\bibfnamefont
  {A.}~\bibnamefont {Laing}}, \ and\ \bibinfo {author} {\bibfnamefont {M.~G.}\
  \bibnamefont {Thompson}},\ }\href {\doibase 10.1126/science.aar7053}
  {\bibfield  {journal} {\bibinfo  {journal} {Science}\ }\textbf {\bibinfo
  {volume} {360}},\ \bibinfo {pages} {285} (\bibinfo {year}
  {2018})}\BibitemShut {NoStop}%
\bibitem [{\citenamefont {Taballione}\ \emph {et~al.}(2019)\citenamefont
  {Taballione}, \citenamefont {Wolterink}, \citenamefont {Lugani},
  \citenamefont {Eckstein}, \citenamefont {Bell}, \citenamefont {Grootjans},
  \citenamefont {Visscher}, \citenamefont {Geskus}, \citenamefont {Roeloffzen},
  \citenamefont {Renema}, \citenamefont {Walmsley}, \citenamefont {Pinkse},\
  and\ \citenamefont {Boller}}]{Taballione19}%
  \BibitemOpen
  \bibfield  {author} {\bibinfo {author} {\bibfnamefont {C.}~\bibnamefont
  {Taballione}}, \bibinfo {author} {\bibfnamefont {T.~A.~W.}\ \bibnamefont
  {Wolterink}}, \bibinfo {author} {\bibfnamefont {J.}~\bibnamefont {Lugani}},
  \bibinfo {author} {\bibfnamefont {A.}~\bibnamefont {Eckstein}}, \bibinfo
  {author} {\bibfnamefont {B.~A.}\ \bibnamefont {Bell}}, \bibinfo {author}
  {\bibfnamefont {R.}~\bibnamefont {Grootjans}}, \bibinfo {author}
  {\bibfnamefont {I.}~\bibnamefont {Visscher}}, \bibinfo {author}
  {\bibfnamefont {D.}~\bibnamefont {Geskus}}, \bibinfo {author} {\bibfnamefont
  {C.~G.~H.}\ \bibnamefont {Roeloffzen}}, \bibinfo {author} {\bibfnamefont
  {J.~J.}\ \bibnamefont {Renema}}, \bibinfo {author} {\bibfnamefont {I.~A.}\
  \bibnamefont {Walmsley}}, \bibinfo {author} {\bibfnamefont {P.~W.~H.}\
  \bibnamefont {Pinkse}}, \ and\ \bibinfo {author} {\bibfnamefont {K.-J.}\
  \bibnamefont {Boller}},\ }\href {\doibase 10.1364/OE.27.026842} {\bibfield
  {journal} {\bibinfo  {journal} {Opt. Express}\ }\textbf {\bibinfo {volume}
  {27}},\ \bibinfo {pages} {26842} (\bibinfo {year} {2019})}\BibitemShut
  {NoStop}%
\bibitem [{\citenamefont {Sparrow}\ \emph {et~al.}(2018)\citenamefont
  {Sparrow}, \citenamefont {Mart{\'\i}n-L{\'o}pez}, \citenamefont {Maraviglia},
  \citenamefont {Neville}, \citenamefont {Harrold}, \citenamefont {Carolan},
  \citenamefont {Joglekar}, \citenamefont {Hashimoto}, \citenamefont {Matsuda},
  \citenamefont {O’Brien}, \citenamefont {Tew},\ and\ \citenamefont
  {Laing}}]{Sparrow18}%
  \BibitemOpen
  \bibfield  {author} {\bibinfo {author} {\bibfnamefont {C.}~\bibnamefont
  {Sparrow}}, \bibinfo {author} {\bibfnamefont {E.}~\bibnamefont
  {Mart{\'\i}n-L{\'o}pez}}, \bibinfo {author} {\bibfnamefont {N.}~\bibnamefont
  {Maraviglia}}, \bibinfo {author} {\bibfnamefont {A.}~\bibnamefont {Neville}},
  \bibinfo {author} {\bibfnamefont {C.}~\bibnamefont {Harrold}}, \bibinfo
  {author} {\bibfnamefont {J.}~\bibnamefont {Carolan}}, \bibinfo {author}
  {\bibfnamefont {Y.~N.}\ \bibnamefont {Joglekar}}, \bibinfo {author}
  {\bibfnamefont {T.}~\bibnamefont {Hashimoto}}, \bibinfo {author}
  {\bibfnamefont {N.}~\bibnamefont {Matsuda}}, \bibinfo {author} {\bibfnamefont
  {J.~L.}\ \bibnamefont {O’Brien}}, \bibinfo {author} {\bibfnamefont {D.~P.}\
  \bibnamefont {Tew}}, \ and\ \bibinfo {author} {\bibfnamefont
  {A.}~\bibnamefont {Laing}},\ }\href {\doibase 10.1038/s41586-018-0152-9}
  {\bibfield  {journal} {\bibinfo  {journal} {Nature}\ }\textbf {\bibinfo
  {volume} {557}},\ \bibinfo {pages} {660–667} (\bibinfo {year}
  {2018})}\BibitemShut {NoStop}%
\bibitem [{\citenamefont {Harris}\ \emph {et~al.}(2017)\citenamefont {Harris},
  \citenamefont {Steinbrecher}, \citenamefont {Prabhu}, \citenamefont {Lahini},
  \citenamefont {Mower}, \citenamefont {Bunandar}, \citenamefont {Chen},
  \citenamefont {Wong}, \citenamefont {Baehr-Jones}, \citenamefont {Hochberg},
  \citenamefont {Lloyd},\ and\ \citenamefont {Englund}}]{Harris17}%
  \BibitemOpen
  \bibfield  {author} {\bibinfo {author} {\bibfnamefont {N.~C.}\ \bibnamefont
  {Harris}}, \bibinfo {author} {\bibfnamefont {G.~R.}\ \bibnamefont
  {Steinbrecher}}, \bibinfo {author} {\bibfnamefont {M.}~\bibnamefont
  {Prabhu}}, \bibinfo {author} {\bibfnamefont {Y.}~\bibnamefont {Lahini}},
  \bibinfo {author} {\bibfnamefont {J.}~\bibnamefont {Mower}}, \bibinfo
  {author} {\bibfnamefont {D.}~\bibnamefont {Bunandar}}, \bibinfo {author}
  {\bibfnamefont {C.}~\bibnamefont {Chen}}, \bibinfo {author} {\bibfnamefont
  {F.~N.~C.}\ \bibnamefont {Wong}}, \bibinfo {author} {\bibfnamefont
  {T.}~\bibnamefont {Baehr-Jones}}, \bibinfo {author} {\bibfnamefont
  {M.}~\bibnamefont {Hochberg}}, \bibinfo {author} {\bibfnamefont
  {S.}~\bibnamefont {Lloyd}}, \ and\ \bibinfo {author} {\bibfnamefont
  {D.}~\bibnamefont {Englund}},\ }\href {\doibase 10.1038/nphoton.2017.95}
  {\bibfield  {journal} {\bibinfo  {journal} {Nature Photonics}\ }\textbf
  {\bibinfo {volume} {11}},\ \bibinfo {pages} {447–452} (\bibinfo {year}
  {2017})}\BibitemShut {NoStop}%
\bibitem [{\citenamefont {Shen}\ \emph {et~al.}(2017)\citenamefont {Shen},
  \citenamefont {Harris}, \citenamefont {Skirlo}, \citenamefont {Prabhu},
  \citenamefont {Baehr-Jones}, \citenamefont {Hochberg}, \citenamefont {Sun},
  \citenamefont {Zhao}, \citenamefont {Larochelle}, \citenamefont {Englund},\
  and\ \citenamefont {Solja{\v c}i{\'c}}}]{Shen17}%
  \BibitemOpen
  \bibfield  {author} {\bibinfo {author} {\bibfnamefont {Y.}~\bibnamefont
  {Shen}}, \bibinfo {author} {\bibfnamefont {N.~C.}\ \bibnamefont {Harris}},
  \bibinfo {author} {\bibfnamefont {S.}~\bibnamefont {Skirlo}}, \bibinfo
  {author} {\bibfnamefont {M.}~\bibnamefont {Prabhu}}, \bibinfo {author}
  {\bibfnamefont {T.}~\bibnamefont {Baehr-Jones}}, \bibinfo {author}
  {\bibfnamefont {M.}~\bibnamefont {Hochberg}}, \bibinfo {author}
  {\bibfnamefont {X.}~\bibnamefont {Sun}}, \bibinfo {author} {\bibfnamefont
  {S.}~\bibnamefont {Zhao}}, \bibinfo {author} {\bibfnamefont {H.}~\bibnamefont
  {Larochelle}}, \bibinfo {author} {\bibfnamefont {D.}~\bibnamefont {Englund}},
  \ and\ \bibinfo {author} {\bibfnamefont {M.}~\bibnamefont {Solja{\v
  c}i{\'c}}},\ }\href {\doibase 10.1038/nphoton.2017.93} {\bibfield  {journal}
  {\bibinfo  {journal} {Nature Photonics}\ }\textbf {\bibinfo {volume} {11}},\
  \bibinfo {pages} {441–446} (\bibinfo {year} {2017})}\BibitemShut {NoStop}%
\bibitem [{\citenamefont {Zhuang}\ \emph {et~al.}(2015)\citenamefont {Zhuang},
  \citenamefont {Roeloffzen}, \citenamefont {Hoekman}, \citenamefont {Boller},\
  and\ \citenamefont {Lowery}}]{Zhuang15}%
  \BibitemOpen
  \bibfield  {author} {\bibinfo {author} {\bibfnamefont {L.}~\bibnamefont
  {Zhuang}}, \bibinfo {author} {\bibfnamefont {C.~G.~H.}\ \bibnamefont
  {Roeloffzen}}, \bibinfo {author} {\bibfnamefont {M.}~\bibnamefont {Hoekman}},
  \bibinfo {author} {\bibfnamefont {K.-J.}\ \bibnamefont {Boller}}, \ and\
  \bibinfo {author} {\bibfnamefont {A.~J.}\ \bibnamefont {Lowery}},\ }\href
  {\doibase 10.1364/OPTICA.2.000854} {\bibfield  {journal} {\bibinfo  {journal}
  {Optica}\ }\textbf {\bibinfo {volume} {2}},\ \bibinfo {pages} {854} (\bibinfo
  {year} {2015})}\BibitemShut {NoStop}%
\bibitem [{\citenamefont {P{\'e}rez}\ \emph {et~al.}(2017)\citenamefont
  {P{\'e}rez}, \citenamefont {Gasulla}, \citenamefont {Crudgington},
  \citenamefont {Thomson}, \citenamefont {Khokhar}, \citenamefont {Li},
  \citenamefont {Cao}, \citenamefont {Z.},\ and\ \citenamefont
  {Capmany}}]{Perez17}%
  \BibitemOpen
  \bibfield  {author} {\bibinfo {author} {\bibfnamefont {D.}~\bibnamefont
  {P{\'e}rez}}, \bibinfo {author} {\bibfnamefont {I.}~\bibnamefont {Gasulla}},
  \bibinfo {author} {\bibfnamefont {L.}~\bibnamefont {Crudgington}}, \bibinfo
  {author} {\bibfnamefont {D.~J.}\ \bibnamefont {Thomson}}, \bibinfo {author}
  {\bibfnamefont {A.~Z.}\ \bibnamefont {Khokhar}}, \bibinfo {author}
  {\bibfnamefont {K.}~\bibnamefont {Li}}, \bibinfo {author} {\bibfnamefont
  {W.}~\bibnamefont {Cao}}, \bibinfo {author} {\bibfnamefont {M.~G.}\
  \bibnamefont {Z.}}, \ and\ \bibinfo {author} {\bibfnamefont {J.}~\bibnamefont
  {Capmany}},\ }\href {\doibase 10.1038/s41467-017-00714-1} {\bibfield
  {journal} {\bibinfo  {journal} {Nature Communications}\ }\textbf {\bibinfo
  {volume} {8}} (\bibinfo {year} {2017}),\
  10.1038/s41467-017-00714-1}\BibitemShut {NoStop}%
\bibitem [{\citenamefont {Reck}\ \emph {et~al.}(1994)\citenamefont {Reck},
  \citenamefont {Zeilinger}, \citenamefont {Bernstein},\ and\ \citenamefont
  {Bertani}}]{Reck94}%
  \BibitemOpen
  \bibfield  {author} {\bibinfo {author} {\bibfnamefont {M.}~\bibnamefont
  {Reck}}, \bibinfo {author} {\bibfnamefont {A.}~\bibnamefont {Zeilinger}},
  \bibinfo {author} {\bibfnamefont {H.~J.}\ \bibnamefont {Bernstein}}, \ and\
  \bibinfo {author} {\bibfnamefont {P.}~\bibnamefont {Bertani}},\ }\href
  {\doibase 10.1103/PhysRevLett.73.58} {\bibfield  {journal} {\bibinfo
  {journal} {Phys. Rev. Lett.}\ }\textbf {\bibinfo {volume} {73}},\ \bibinfo
  {pages} {58} (\bibinfo {year} {1994})}\BibitemShut {NoStop}%
\bibitem [{\citenamefont {Clements}\ \emph {et~al.}(2016)\citenamefont
  {Clements}, \citenamefont {Humphreys}, \citenamefont {Metcalf}, \citenamefont
  {Kolthammer},\ and\ \citenamefont {Walmsley}}]{Clements16}%
  \BibitemOpen
  \bibfield  {author} {\bibinfo {author} {\bibfnamefont {W.~R.}\ \bibnamefont
  {Clements}}, \bibinfo {author} {\bibfnamefont {P.~C.}\ \bibnamefont
  {Humphreys}}, \bibinfo {author} {\bibfnamefont {B.~J.}\ \bibnamefont
  {Metcalf}}, \bibinfo {author} {\bibfnamefont {W.~S.}\ \bibnamefont
  {Kolthammer}}, \ and\ \bibinfo {author} {\bibfnamefont {I.~A.}\ \bibnamefont
  {Walmsley}},\ }\href {\doibase 10.1364/OPTICA.3.001460} {\bibfield  {journal}
  {\bibinfo  {journal} {Optica}\ }\textbf {\bibinfo {volume} {3}},\ \bibinfo
  {pages} {1460} (\bibinfo {year} {2016})}\BibitemShut {NoStop}%
\bibitem [{\citenamefont {Miller}(2015)}]{Miller15}%
  \BibitemOpen
  \bibfield  {author} {\bibinfo {author} {\bibfnamefont {D.~A.~B.}\
  \bibnamefont {Miller}},\ }\href {\doibase 10.1364/OPTICA.2.000747} {\bibfield
   {journal} {\bibinfo  {journal} {Optica}\ }\textbf {\bibinfo {volume} {2}},\
  \bibinfo {pages} {747} (\bibinfo {year} {2015})}\BibitemShut {NoStop}%
\bibitem [{\citenamefont {Ribeiro}\ \emph {et~al.}(2016)\citenamefont
  {Ribeiro}, \citenamefont {Ruocco}, \citenamefont {Vanacker},\ and\
  \citenamefont {Bogaerts}}]{Ribeiro16}%
  \BibitemOpen
  \bibfield  {author} {\bibinfo {author} {\bibfnamefont {A.}~\bibnamefont
  {Ribeiro}}, \bibinfo {author} {\bibfnamefont {A.}~\bibnamefont {Ruocco}},
  \bibinfo {author} {\bibfnamefont {L.}~\bibnamefont {Vanacker}}, \ and\
  \bibinfo {author} {\bibfnamefont {W.}~\bibnamefont {Bogaerts}},\ }\href
  {\doibase 10.1364/OPTICA.3.001348} {\bibfield  {journal} {\bibinfo  {journal}
  {Optica}\ }\textbf {\bibinfo {volume} {3}},\ \bibinfo {pages} {1348}
  (\bibinfo {year} {2016})}\BibitemShut {NoStop}%
\bibitem [{\citenamefont {Fldzhyan}\ \emph {et~al.}(2020)\citenamefont
  {Fldzhyan}, \citenamefont {Saygin},\ and\ \citenamefont
  {Kulik}}]{Fldzhyan20}%
  \BibitemOpen
  \bibfield  {author} {\bibinfo {author} {\bibfnamefont {S.~A.}\ \bibnamefont
  {Fldzhyan}}, \bibinfo {author} {\bibfnamefont {M.~Y.}\ \bibnamefont
  {Saygin}}, \ and\ \bibinfo {author} {\bibfnamefont {S.~P.}\ \bibnamefont
  {Kulik}},\ }\href {\doibase 10.1364/OL.385433} {\bibfield  {journal}
  {\bibinfo  {journal} {Opt. Lett.}\ }\textbf {\bibinfo {volume} {45}},\
  \bibinfo {pages} {2632} (\bibinfo {year} {2020})}\BibitemShut {NoStop}%
\bibitem [{\citenamefont {Burgwal}\ \emph {et~al.}(2017)\citenamefont
  {Burgwal}, \citenamefont {Clements}, \citenamefont {Smith}, \citenamefont
  {Gates}, \citenamefont {Kolthammer}, \citenamefont {Renema},\ and\
  \citenamefont {Walmsley}}]{Burgwal17}%
  \BibitemOpen
  \bibfield  {author} {\bibinfo {author} {\bibfnamefont {R.}~\bibnamefont
  {Burgwal}}, \bibinfo {author} {\bibfnamefont {W.~R.}\ \bibnamefont
  {Clements}}, \bibinfo {author} {\bibfnamefont {D.~H.}\ \bibnamefont {Smith}},
  \bibinfo {author} {\bibfnamefont {J.~C.}\ \bibnamefont {Gates}}, \bibinfo
  {author} {\bibfnamefont {W.~S.}\ \bibnamefont {Kolthammer}}, \bibinfo
  {author} {\bibfnamefont {J.~J.}\ \bibnamefont {Renema}}, \ and\ \bibinfo
  {author} {\bibfnamefont {I.~A.}\ \bibnamefont {Walmsley}},\ }\href {\doibase
  10.1364/OE.25.028236} {\bibfield  {journal} {\bibinfo  {journal} {Opt.
  Express}\ }\textbf {\bibinfo {volume} {25}},\ \bibinfo {pages} {28236}
  (\bibinfo {year} {2017})}\BibitemShut {NoStop}%
\bibitem [{\citenamefont {Pai}\ \emph {et~al.}(2019)\citenamefont {Pai},
  \citenamefont {Bartlett}, \citenamefont {Solgaard},\ and\ \citenamefont
  {Miller}}]{Pai19}%
  \BibitemOpen
  \bibfield  {author} {\bibinfo {author} {\bibfnamefont {S.}~\bibnamefont
  {Pai}}, \bibinfo {author} {\bibfnamefont {B.}~\bibnamefont {Bartlett}},
  \bibinfo {author} {\bibfnamefont {O.}~\bibnamefont {Solgaard}}, \ and\
  \bibinfo {author} {\bibfnamefont {D.~A.~B.}\ \bibnamefont {Miller}},\ }\href
  {\doibase 10.1103/PhysRevApplied.11.064044} {\bibfield  {journal} {\bibinfo
  {journal} {Phys. Rev. Applied}\ }\textbf {\bibinfo {volume} {11}},\ \bibinfo
  {pages} {064044} (\bibinfo {year} {2019})}\BibitemShut {NoStop}%
\bibitem [{\citenamefont {Mennea}\ \emph {et~al.}(2018)\citenamefont {Mennea},
  \citenamefont {Clements}, \citenamefont {Smith}, \citenamefont {Gates},
  \citenamefont {Metcalf}, \citenamefont {Bannerman}, \citenamefont {Burgwal},
  \citenamefont {Renema}, \citenamefont {Kolthammer}, \citenamefont
  {Walmsley},\ and\ \citenamefont {Smith}}]{Mennea18}%
  \BibitemOpen
  \bibfield  {author} {\bibinfo {author} {\bibfnamefont {P.~L.}\ \bibnamefont
  {Mennea}}, \bibinfo {author} {\bibfnamefont {W.~R.}\ \bibnamefont
  {Clements}}, \bibinfo {author} {\bibfnamefont {D.~H.}\ \bibnamefont {Smith}},
  \bibinfo {author} {\bibfnamefont {J.~C.}\ \bibnamefont {Gates}}, \bibinfo
  {author} {\bibfnamefont {B.~J.}\ \bibnamefont {Metcalf}}, \bibinfo {author}
  {\bibfnamefont {R.~H.~S.}\ \bibnamefont {Bannerman}}, \bibinfo {author}
  {\bibfnamefont {R.}~\bibnamefont {Burgwal}}, \bibinfo {author} {\bibfnamefont
  {J.~J.}\ \bibnamefont {Renema}}, \bibinfo {author} {\bibfnamefont {W.~S.}\
  \bibnamefont {Kolthammer}}, \bibinfo {author} {\bibfnamefont {I.~A.}\
  \bibnamefont {Walmsley}}, \ and\ \bibinfo {author} {\bibfnamefont {P.~G.~R.}\
  \bibnamefont {Smith}},\ }\href {\doibase 10.1364/OPTICA.5.001087} {\bibfield
  {journal} {\bibinfo  {journal} {Optica}\ }\textbf {\bibinfo {volume} {5}},\
  \bibinfo {pages} {1087} (\bibinfo {year} {2018})}\BibitemShut {NoStop}%
\bibitem [{\citenamefont {Boes}\ \emph {et~al.}(2018)\citenamefont {Boes},
  \citenamefont {Corcoran}, \citenamefont {Chang}, \citenamefont {Bowers},\
  and\ \citenamefont {Mitchell}}]{Boes18}%
  \BibitemOpen
  \bibfield  {author} {\bibinfo {author} {\bibfnamefont {A.}~\bibnamefont
  {Boes}}, \bibinfo {author} {\bibfnamefont {B.}~\bibnamefont {Corcoran}},
  \bibinfo {author} {\bibfnamefont {L.}~\bibnamefont {Chang}}, \bibinfo
  {author} {\bibfnamefont {J.}~\bibnamefont {Bowers}}, \ and\ \bibinfo {author}
  {\bibfnamefont {A.}~\bibnamefont {Mitchell}},\ }\href {\doibase
  https://doi.org/10.1002/lpor.201700256} {\bibfield  {journal} {\bibinfo
  {journal} {Laser \& Photonics Reviews}\ }\textbf {\bibinfo {volume} {12}},\
  \bibinfo {pages} {1700256} (\bibinfo {year} {2018})}\BibitemShut {NoStop}%
\bibitem [{\citenamefont {Taballione}\ \emph {et~al.}(2020)\citenamefont
  {Taballione}, \citenamefont {van~der Meer}, \citenamefont {Snijders},
  \citenamefont {Hooijschuur}, \citenamefont {Epping}, \citenamefont
  {de~Goede}, \citenamefont {Kassenberg}, \citenamefont {Venderbosch},
  \citenamefont {Toebes}, \citenamefont {van~den Vlekkert}, \citenamefont
  {Pinkse},\ and\ \citenamefont {Renema}}]{Taballione20}%
  \BibitemOpen
  \bibfield  {author} {\bibinfo {author} {\bibfnamefont {C.}~\bibnamefont
  {Taballione}}, \bibinfo {author} {\bibfnamefont {R.}~\bibnamefont {van~der
  Meer}}, \bibinfo {author} {\bibfnamefont {H.}~\bibnamefont {Snijders}},
  \bibinfo {author} {\bibfnamefont {P.}~\bibnamefont {Hooijschuur}}, \bibinfo
  {author} {\bibfnamefont {J.}~\bibnamefont {Epping}}, \bibinfo {author}
  {\bibfnamefont {M.}~\bibnamefont {de~Goede}}, \bibinfo {author}
  {\bibfnamefont {B.}~\bibnamefont {Kassenberg}}, \bibinfo {author}
  {\bibfnamefont {P.}~\bibnamefont {Venderbosch}}, \bibinfo {author}
  {\bibfnamefont {C.}~\bibnamefont {Toebes}}, \bibinfo {author} {\bibfnamefont
  {H.}~\bibnamefont {van~den Vlekkert}}, \bibinfo {author} {\bibfnamefont
  {P.}~\bibnamefont {Pinkse}}, \ and\ \bibinfo {author} {\bibfnamefont
  {J.}~\bibnamefont {Renema}},\ }\href@noop {} {\bibfield  {journal} {\bibinfo
  {journal} {arxiv:2012.05673}\ } (\bibinfo {year} {2020})}\BibitemShut
  {NoStop}%
\bibitem [{\citenamefont {Arrazola}\ \emph {et~al.}(2021)\citenamefont
  {Arrazola}, \citenamefont {Bergholm}, \citenamefont {Brádler}, \citenamefont
  {Bromley}, \citenamefont {Collins}, \citenamefont {Dhand}, \citenamefont
  {Fumagalli}, \citenamefont {Gerrits}, \citenamefont {Goussev}, \citenamefont
  {Helt}, \citenamefont {Hundal}, \citenamefont {Isacsson}, \citenamefont
  {Israel}, \citenamefont {Izaac}, \citenamefont {Jahangiri}, \citenamefont
  {Janik}, \citenamefont {Killoran}, \citenamefont {Kumar}, \citenamefont
  {Lavoie}, \citenamefont {Lita}, \citenamefont {Mahler}, \citenamefont
  {Menotti}, \citenamefont {Morrison}, \citenamefont {Nam}, \citenamefont
  {Neuhaus}, \citenamefont {Qi}, \citenamefont {Quesada}, \citenamefont
  {Repingon}, \citenamefont {Sabapathy}, \citenamefont {Schuld}, \citenamefont
  {Su}, \citenamefont {Swinarton}, \citenamefont {Száva}, \citenamefont {Tan},
  \citenamefont {Tan}, \citenamefont {Vaidya}, \citenamefont {Vernon},
  \citenamefont {Zabaneh},\ and\ \citenamefont {Zhang}}]{Arrazola21}%
  \BibitemOpen
  \bibfield  {author} {\bibinfo {author} {\bibfnamefont {J.}~\bibnamefont
  {Arrazola}}, \bibinfo {author} {\bibfnamefont {V.}~\bibnamefont {Bergholm}},
  \bibinfo {author} {\bibfnamefont {K.}~\bibnamefont {Brádler}}, \bibinfo
  {author} {\bibfnamefont {T.~R.}\ \bibnamefont {Bromley}}, \bibinfo {author}
  {\bibfnamefont {M.~J.}\ \bibnamefont {Collins}}, \bibinfo {author}
  {\bibfnamefont {I.}~\bibnamefont {Dhand}}, \bibinfo {author} {\bibfnamefont
  {A.}~\bibnamefont {Fumagalli}}, \bibinfo {author} {\bibfnamefont
  {T.}~\bibnamefont {Gerrits}}, \bibinfo {author} {\bibfnamefont
  {A.}~\bibnamefont {Goussev}}, \bibinfo {author} {\bibfnamefont {L.~G.}\
  \bibnamefont {Helt}}, \bibinfo {author} {\bibfnamefont {J.}~\bibnamefont
  {Hundal}}, \bibinfo {author} {\bibfnamefont {T.}~\bibnamefont {Isacsson}},
  \bibinfo {author} {\bibfnamefont {R.~B.}\ \bibnamefont {Israel}}, \bibinfo
  {author} {\bibfnamefont {J.}~\bibnamefont {Izaac}}, \bibinfo {author}
  {\bibfnamefont {S.}~\bibnamefont {Jahangiri}}, \bibinfo {author}
  {\bibfnamefont {R.}~\bibnamefont {Janik}}, \bibinfo {author} {\bibfnamefont
  {N.}~\bibnamefont {Killoran}}, \bibinfo {author} {\bibfnamefont {S.~P.}\
  \bibnamefont {Kumar}}, \bibinfo {author} {\bibfnamefont {J.}~\bibnamefont
  {Lavoie}}, \bibinfo {author} {\bibfnamefont {A.~E.}\ \bibnamefont {Lita}},
  \bibinfo {author} {\bibfnamefont {D.~H.}\ \bibnamefont {Mahler}}, \bibinfo
  {author} {\bibfnamefont {M.}~\bibnamefont {Menotti}}, \bibinfo {author}
  {\bibfnamefont {B.}~\bibnamefont {Morrison}}, \bibinfo {author}
  {\bibfnamefont {S.~W.}\ \bibnamefont {Nam}}, \bibinfo {author} {\bibfnamefont
  {L.}~\bibnamefont {Neuhaus}}, \bibinfo {author} {\bibfnamefont {H.~Y.}\
  \bibnamefont {Qi}}, \bibinfo {author} {\bibfnamefont {N.}~\bibnamefont
  {Quesada}}, \bibinfo {author} {\bibfnamefont {A.}~\bibnamefont {Repingon}},
  \bibinfo {author} {\bibfnamefont {K.~K.}\ \bibnamefont {Sabapathy}}, \bibinfo
  {author} {\bibfnamefont {M.}~\bibnamefont {Schuld}}, \bibinfo {author}
  {\bibfnamefont {D.}~\bibnamefont {Su}}, \bibinfo {author} {\bibfnamefont
  {J.}~\bibnamefont {Swinarton}}, \bibinfo {author} {\bibfnamefont
  {A.}~\bibnamefont {Száva}}, \bibinfo {author} {\bibfnamefont
  {K.}~\bibnamefont {Tan}}, \bibinfo {author} {\bibfnamefont {P.}~\bibnamefont
  {Tan}}, \bibinfo {author} {\bibfnamefont {V.~D.}\ \bibnamefont {Vaidya}},
  \bibinfo {author} {\bibfnamefont {Z.}~\bibnamefont {Vernon}}, \bibinfo
  {author} {\bibfnamefont {Z.}~\bibnamefont {Zabaneh}}, \ and\ \bibinfo
  {author} {\bibfnamefont {Y.}~\bibnamefont {Zhang}},\ }\href {\doibase
  10.1038/s41586-021-03202-1} {\bibfield  {journal} {\bibinfo  {journal}
  {Nature}\ }\textbf {\bibinfo {volume} {591}},\ \bibinfo {pages} {54}
  (\bibinfo {year} {2021})}\BibitemShut {NoStop}%
\end{thebibliography}%

\end{document}